\documentclass[aps,pra,reprint,superscriptaddress,amsmath,amssymb,longbibliography]{revtex4-2}

\usepackage{graphicx}
\usepackage{longtable}
\usepackage{xcolor}
\usepackage[bookma rks,bookmarksnumbered,colorlinks,allcolors=blue]{hyperref}
\usepackage{bookmark}
\usepackage{ifthen}
\usepackage{ulem}
\usepackage{physics}
\usepackage{makecell}
\usepackage{chngcntr}
\usepackage{appendix}
\usepackage{hyperref}

\definecolor{dkgreen}{rgb}{0,0.6,0}
\definecolor{gray}{rgb}{0.5,0.5,0.5}
\definecolor{mauve}{rgb}{0.58,0,0.82}

\def\be#1\ee{\begin{equation}#1\end{equation}}
\def\ba#1\ea{\begin{align}#1\end{align}}
\def\bg#1\eg{\begin{gather}#1\end{gather}}

\newcommand{\Phiext}{\Phi_{\text{ext}}}

\begin{document}

\title{High coherence fluxonium manufactured with a wafer-scale uniformity process}

\author{Fei Wang}
\thanks{These authors contributed equally to this work}
\affiliation{Quantum Science Center of Guangdong-Hong Kong-Macao Greater Bay Area, Shenzhen 518045, China}
\affiliation{Z-Axis Quantum}
\author{Kannan Lu}
\thanks{These authors contributed equally to this work}
\affiliation{Quantum Science Center of Guangdong-Hong Kong-Macao Greater Bay Area, Shenzhen 518045, China}
\affiliation{Z-Axis Quantum}
\author{Huijuan Zhan}
\affiliation{Quantum Science Center of Guangdong-Hong Kong-Macao Greater Bay Area, Shenzhen 518045, China}
\author{Lu Ma}
\affiliation{Quantum Science Center of Guangdong-Hong Kong-Macao Greater Bay Area, Shenzhen 518045, China}
\affiliation{Z-Axis Quantum}
\author{Feng Wu}
\affiliation{Zhongguancun Laboratory, Beijing, China}
\author{Hantao Sun}
\affiliation{China Telecom Quantum Information Technology Group Co., Ltd., Hefei 230031, China}
\author{Hao Deng}
\affiliation{International Center for Quantum Materials, Peking University, Beijing, 100871, China}
\author{Yang Bai}
\affiliation{MetaPhotons Co., Ltd., Hangzhou, China}
\author{Feng Bao}
\affiliation{Institute of Quantum Sensing, Zhejiang University, Hangzhou 310007, China}
\author{Xu Chang}
\affiliation{Shanghai E-Matterwave Sci \& Tech Co., Ltd., Shanghai 201100, China}
\author{Ran Gao}
\affiliation{Quantum Science Center of Guangdong-Hong Kong-Macao Greater Bay Area, Shenzhen 518045, China}
\affiliation{Z-Axis Quantum}
\author{Xun Gao}
\affiliation{Kirkland, WA 98033, USA}
\author{Guicheng Gong}
\affiliation{Quantum Science Center of Guangdong-Hong Kong-Macao Greater Bay Area, Shenzhen 518045, China}
\affiliation{Z-Axis Quantum}
\author{Lijuan Hu}
\affiliation{Quantum Science Center of Guangdong-Hong Kong-Macao Greater Bay Area, Shenzhen 518045, China}
\author{Ruizi Hu}
\affiliation{Quantum Science Center of Guangdong-Hong Kong-Macao Greater Bay Area, Shenzhen 518045, China}
\author{Honghong Ji}
\affiliation{Quantum Science Center of Guangdong-Hong Kong-Macao Greater Bay Area, Shenzhen 518045, China}
\author{Xizheng Ma}
\affiliation{Quantum Science Center of Guangdong-Hong Kong-Macao Greater Bay Area, Shenzhen 518045, China}
\author{Liyong Mao}
\affiliation{Quantum Science Center of Guangdong-Hong Kong-Macao Greater Bay Area, Shenzhen 518045, China}
\affiliation{Z-Axis Quantum}
\author{Zhijun Song}
\affiliation{Shanghai E-Matterwave Sci \& Tech Co., Ltd., Shanghai 201100, China}
\author{Chengchun Tang}
\affiliation{Institute of Quantum Sensing, Zhejiang University, Hangzhou 310007, China}
\author{Hongcheng Wang}
\affiliation{Quantum Science Center of Guangdong-Hong Kong-Macao Greater Bay Area, Shenzhen 518045, China}
\author{Tenghui Wang}
\affiliation{Quantum Science Center of Guangdong-Hong Kong-Macao Greater Bay Area, Shenzhen 518045, China}
\author{Ziang Wang}
\affiliation{Quantum Science Center of Guangdong-Hong Kong-Macao Greater Bay Area, Shenzhen 518045, China}
\affiliation{Zhejiang Institute of Modern Physics, Zhejiang University, Hangzhou 310027, China}
\author{Tian Xia}
\affiliation{Huaxin Jushu Microelectronics Co., Ltd., Hangzhou, China}
\author{Hongxin Xu}
\affiliation{Hefei Everacq Technology Co., Ltd., Heifei, China}
\author{Ze Zhan}
\affiliation{Quantum Science Center of Guangdong-Hong Kong-Macao Greater Bay Area, Shenzhen 518045, China}
\affiliation{Z-Axis Quantum}
\author{Gengyan Zhang}
\affiliation{Zhejiang Laboratory, Hangzhou, China}
\author{Tao Zhou}
\affiliation{Quantum Science Center of Guangdong-Hong Kong-Macao Greater Bay Area, Shenzhen 518045, China}
\author{Mengyu Zhu}
\affiliation{Quantum Science Center of Guangdong-Hong Kong-Macao Greater Bay Area, Shenzhen 518045, China}
\affiliation{Z-Axis Quantum}
\author{Qingbin Zhu}
\affiliation{Shanghai E-Matterwave Sci \& Tech Co., Ltd., Shanghai 201100, China}
\author{Shasha Zhu}
\affiliation{Quantum Science Center of Guangdong-Hong Kong-Macao Greater Bay Area, Shenzhen 518045, China}
\affiliation{Z-Axis Quantum}
\author{Xing Zhu}
\affiliation{Quantum Science Center of Guangdong-Hong Kong-Macao Greater Bay Area, Shenzhen 518045, China}
\affiliation{Z-Axis Quantum}
\author{Yaoyun Shi}
\affiliation{Z-Axis Quantum}
\author{Hui-Hai Zhao}
\affiliation{Zhongguancun Laboratory, Beijing, China}
\author{Chunqing Deng}
\email{dengchunqing@quantumsc.cn}
\affiliation{Quantum Science Center of Guangdong-Hong Kong-Macao Greater Bay Area, Shenzhen 518045, China}
\affiliation{Z-Axis Quantum}

\begin{abstract}
Fluxonium qubits are recognized for their high coherence times and high operation fidelities, attributed to their unique design incorporating a superinductor, which is typically implemented using an array of over 100 Josephson junctions.
However, this complexity poses significant fabrication challenges, particularly in achieving high yield and junction uniformity with traditional methods. Here, we introduce an overlap process for Josephson junction fabrication 
that achieves nearly 100\% yield and maintains uniformity across a 2-inch wafer with less than 5\% variation for the phase slip junction and less than 2\% for the entire junction array.
We use a compact junction array design that achieves state-of-the-art dielectric loss tangents and flux noise levels, as confirmed by multiple devices. This enables fluxonium qubits to reach energy relaxation times exceeding 1 millisecond at the flux frustration point.
This work paves the way for scalable high coherence fluxonium processors using CMOS-compatible processes, marking a significant step towards practical quantum computing.
\end{abstract}
\maketitle

\bookmarksetup{startatroot}

\section{Introduction}

Fluxonium qubits~\cite{manucharyan2009fluxonium} have garnered increasing interest owing to their significantly longer coherence times~\cite{nguyen2019high, zhang2021universal, somoroff2021millisecond} and enhanced fidelity in gate operations~\cite{bao2022fluxonium, ma2024native, zhang2023tunable, ding2023highfidelity}, positioning them as a viable alternative for the development of fault-tolerant superconducting quantum processors. This superior performance is attributed to the intrinsic characteristics of their energy spectrum. In particular, the fluxonium spectrum exhibits a first-order insensitivity to external flux variations at the flux frustration point, effectively creating a dephasing-resistant sweet spot. At this juncture, the qubit's energy gap $f_{01}$ is determined by the quantum tunneling between two fluxon states within a double-well potential. This tunneling, known as coherent quantum phase slip~\cite{astafiev2012coherent}, has its amplitude exponentially suppressed by the Josephson energy barrier of the small phase slip junction. As a result, fluxonium qubits operate at significantly lower frequencies ($f_{01} < 1$~GHz) compared to transmon qubits~\cite{koch2007charge}. Consequently, this operational frequency reduction bolsters the qubits' $T_1$ lifetime, assuming a similar quality factor $Q = 2\pi f_{01}T_1$. Critically, the diminution in qubit frequency does not translate to a compromise in operational speed. The inherent substantial anharmonicity of the fluxonium spectrum has inspired a variety of proposals for executing high-fidelity operations~\cite{nesterov2018microwave, moskalenko2021tunable, nguyen2022blueprint, nesterov2022cnot, weiss2022high, simakov2024high}. Building on these proposals, significant advancements have been made in implementing rapid and high-fidelity two-qubit gates~\cite{ficheux2021fast, bao2022fluxonium, moskalenko2022high, zhang2023tunable, ding2023highfidelity, simakov2023coupler, ma2024native}, as well as in readout~\cite{gusenkova2021quantum} and reset techniques~\cite{gebauer2020state, wang2024efficient}.

The advantageous characteristics of fluxonium qubits are offset by their increased fabrication complexity relative to transmon qubits. The simplicity of transmon qubits allows for the widespread adoption of double-angle evaporation techniques to construct Josephson junctions (JJ) with relatively reliable process control and accurate parameter targeting~\cite{Kreikebaum2020,Takahashi2023,Muth2023,Pishchimova2023,Moskalev2023}. In contrast, the fabrication of each fluxonium qubit not only involves creating a phase slip JJ (orange box in Fig.~\ref{fig:sample}b) but also incorporating an array of approximately 100 JJs (green box in Fig.~\ref{fig:sample}b) to function as a large shunting inductor. This complexity introduces substantial challenges for device yield through increased variability in qubit parameters. Two primary issues exacerbate parameter variability in fluxonium qubit fabrication: the impact of film surface and pattern edge roughness, as well as sidewall deposition effects due to angle evaporation, both of which constrain the accuracy of phase slip JJ parameters similar to transmons. Additionally, the effective junction area in Josephson junction arrays (JJAs) is highly sensitive to the evaporation angle, which varies significantly across the wafer surface, complicating consistent fabrication on larger wafers.

To address this limitation and facilitate scalable fluxonium processor production, we developed an overlap junction process for fluxonium qubit fabrication~\cite{Wu2017, stehli2020josephson, Verjauw2022}. In the overlap junction process flow, no wafer tilting is required for junction electrodes deposition, and the top and bottom electrode formations are performed in two separated patterning and deposition steps, with a vacuum break in between. This overlap approach has exhibited nearly 100\% yield and uniformity across a 2-inch wafer with less than 5\% variation for the phase slip junction and less than 2\% for the junction array. The variation in the JJA is characterized by the variation in the total normal resistance $R_n$ of all the junctions in series within the entire array. Across multiple fluxonium devices, we reached dielectric loss tangent $\tan\delta_{C} \in (1.2, 5.0)\times 10^{-6}$ and $1/f$ flux noise amplitude $A_\Phi \in (1.4, 2.6)~\mu\Phi_{0}/\sqrt{\text{Hz}}$; here $\Phi_{0} = h/(2e)$ is the flux quantum. On our best device, we achieved $T_{1} = 1.168$~ms and $T_{2, \mathrm{echo}} = 0.943$~ms at the qubit's flux frustration position with the qubit frequency $f_{01} = 197$~MHz. 
Compared with our previous work~\cite{Sun2023}, we attribute the improvement on coherence times to the flux noise reduction due to the robust compact JJA enabled by our overlap process.
The dielectric loss and flux noise levels are comparable with the best performing fluxonium qubits made with the angle-evaporation techniques and embedded in a 3-dimensional cavity, tailored for high coherence demonstration~\cite{nguyen2019high, somoroff2021millisecond}. Our work suggests the feasibility of scaling high coherence fluxonium processor with a CMOS-compatible fabrication process. 

\section{Fluxonium Qubits with overlap junctions}

\begin{figure}
    \includegraphics[width = 1\columnwidth]{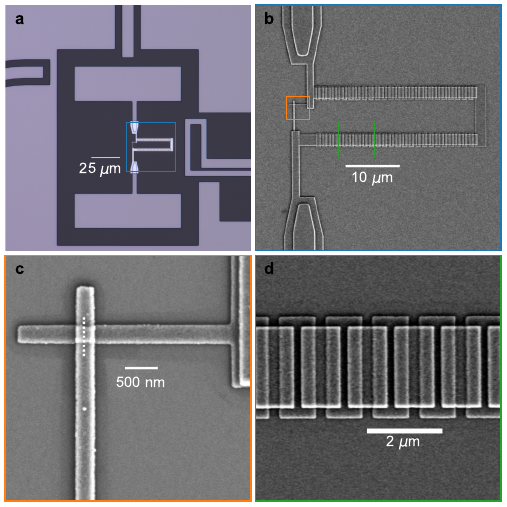}
    \caption{$\bf{a}$ Optical micrograph of a typical fluxonium qubit device coupled to a charge driving line, a flux line, and a readout cavity.
    $\bf{b}$ Scanning electron microscopy (SEM) image of the region enclosed by the blue box in $\bf{a}$, representing the qubit loop structure, composed of a phase slip Josephson junction (JJ) and a Josephson junction array (JJA) $\bf{c}$ SEM image of the phase slip JJ in orange box in $\bf{b}$. The dashed line depicts the cross sectional view in \autoref{fig:junction_Rn}. $\bf{d}$ SEM image of the green box region in $\bf{b}$, focusing on the compact JJA.}
    \label{fig:sample}
\end{figure}

\begin{figure}
    \includegraphics[width = 1\columnwidth]{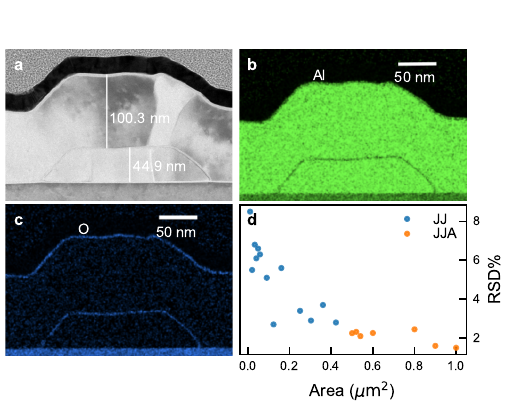}
    \caption{$\bf{a}$ Cross sectional transmission electron microscopy (TEM) image of a phase slip JJ viewed along the direction shown as the dashed line in \autoref{fig:sample}c. $\bf{b}$ and $\bf{c}$ Energy dispersive X-ray spectroscopy (EDX) of the cross section showing Al and O concentrations, respectively. $\bf{d}$ Within wafer relative standard deviation of the junction $R_{n}$ as a function of the junction area for both the phase slip JJ and JJA.
}
    \label{fig:junction_Rn}
\end{figure}

We choose 2-inch sapphire wafer as the substrate for our fluxonium processor due to its low dielectric loss~\cite{read2023precision}. Tantalum film is deposited on the sapphire substrate to form superconducting circuit elements except for the formation of the qubit loop. As shown in \autoref{fig:sample}a-d, our fluxonium qubit consists of a single phase slip Josephson junction (referred as JJ hereafter) connected in parallel with a shunting capacitor and an array of 100 Josephson junctions as the shunting inductor. In this work, we adopt an overlap junction process combined with lift-off technique (see Appendix~\ref{appendix:fabrication}) to fabricate JJ/JJA with Al/AlO$_x$/Al structure. The process starts with an Al deposition to form the bottom electrode (BE), followed by e-beam resist lift-off and top electrode (TE) patterning. The wafer is then loaded into vacuum chamber for in situ barrier AlO$_x$ formation and TE Al deposition. Prior to AlO$_x$ formation, we implement a well calibrated argon (Ar) milling process to remove native oxides for AlO$_x$ barrier quality control. 

To evaluate qubit yield and process variability, we measure normal resistance ($R_{n}$) of JJ and JJA with various junction sizes on 2-inch test structure wafers. The yield is defined as no hard failure (opens/shorts) of qubits on test patterns, since real devices/qubits cannot be measured without any damage. Each test wafer contains 16 dies with 16 JJ or JJA of various sizes on each die.
Excluding a few data points from wafer-edge dies with photolithography issues, almost all of the JJ/JJA structures show expected $R_{n}$ values from wafers we measured, which is an indication of near perfect qubit yield on our device wafers. For the JJs, the across wafer relative standard deviation (RSD\%) of $R_{n}$ trends down with increasing junction size, from 8.5\% at 0.01~$\mu m^{2}$ to 2.8\% at 0.42~$\mu m^{2}$. For JJA with single junction size of 0.5~$\mu m^{2}$ to 1~$\mu m^{2}$, RSD\% is lower at 1.5-2.5\% compared to smaller phase slip JJ, shown in \autoref{fig:junction_Rn}d. We emphasize that the RSD\% of the JJA is measured by the within-wafer variation in the normal resistance $R_n$ of the entire array, which characterizes the variation in the total inductance formed by the JJA. In our fluxonium design, the typical junction dimensions for JJ and JJA are 0.05~$\mu m^{2}$ and 1~$\mu m^{2}$, with RSD\% being 5\% and 1.5\% respectively. Transmission electron microscope (TEM) cross section of a typical single phase slip JJ made with the overlap process is shown in \autoref{fig:junction_Rn}a. The thickness target of BE Al film is 60~nm, and it came out at 45~nm in the final structure, which means the Ar ion milling consumes 15~nm of Al prior to barrier AlO$_x$ formation. The compositional analysis performed with the energy dispersive X-ray spectroscopy (EDX) of the TEM shows mostly Al concentration in the junction with O occurring in a smooth interface between top and bottom electrodes, as depicted in \autoref{fig:junction_Rn}b and c.

It is also worth mentioning that we have been implementing the overlap junction process in our 22-qubit fluxonium processor fabrication on 4-inch sapphire wafers (see Appendix~\ref{appendix:4in} for the statistics of the test structures). At the time of writing, two 22-qubit fluxonium chips from different wafers have been fully measured and characterized, and all qubits show normal operation with relatively accurate qubit parameters. This is strong evidence that the overlap junction process can be easily transferred to even larger wafer scale without any obvious issue. However, the process flow presented in this study, which utilizes e-beam lithography and lift-off techniques for JJ/JJA formation, is not immediately compatible with the contemporary 300~mm semiconductor manufacturing process. Future research will investigate the use of photolithography and subtractive etching for JJ/JJA formation, aiming to achieve a CMOS-compatible process flow for fluxonium processors.

\section{Demonstration of high coherence fluxonium}

\begin{figure}[tb]
    \includegraphics[width = 1\columnwidth]{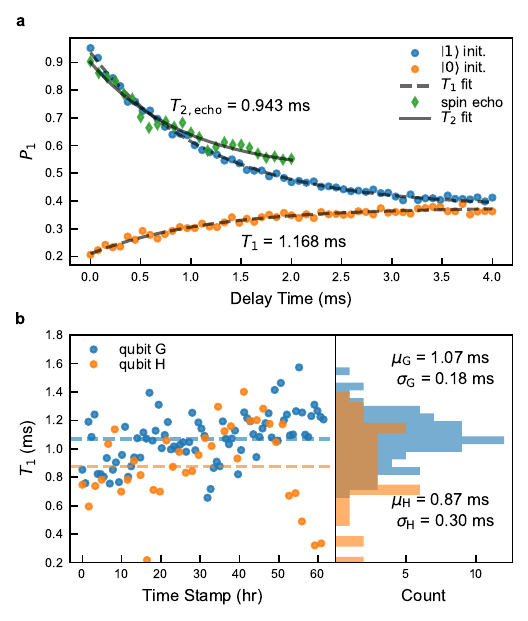}
    \caption{$\bf{a}$ $T_{1}$ (blue and orange) and $T_{2, \mathrm{echo}}$ (green) decay curves for qubit G at the flux frustration position. The $T_{1}$ decays, for both the $\ket{0}$ and $\ket{1}$ initialization (dashed black lines), are simultaneously fitted to two exponential functions with the same decay constant. The $T_{2, \mathrm{echo}}$ decay is fitted with an single exponential function (solid black line). $\bf{b}$ Temporal fluctuation of $T_{1}$ for qubits G (blue) and H (orange), including mean values (dashed lines) and distribution histograms.}
    \label{fig:t1_t2_sweetspot}
\end{figure}

In \autoref{fig:t1_t2_sweetspot}a, we present the measured energy relaxation ($T_{1}$) and dephasing ($T_{2}$) processes of one qubit at its flux frustration position. In the $T_{1}$ measurement, we initialize the qubit in either the ground $\ket{0}$ or excited $\ket{1}$ state at the beginning of each experiment via
projective readout and herald the desired initial state (see Appendix~\ref{appendix:measurement_protocols}). 
The postselected population decay from $\ket{1}$ and $\ket{0}$ is jointly fitted with two exponential decays of the same decay constant $T_{1}$ and thermal state population $b$, i.e. $(P_{1, \ket{1}}, P_{1, \ket{0}}) = (a_{1}\exp(-t/T_{1})+b, a_{2}\exp(-t/T_{1}) + b)$, where $P_{i, \ket{j}}$ means measuring the $\ket{i}$ population with qubit initialized at $\ket{j}$.
We obtain $T_{1} = 1.168$~ms and estimate the qubit's effective temperature $T = 18.7$~mK from the thermal state population. 
In the $T_{2}$ measurement, we utilize the spin echo pulse sequence after qubit initialization. 
For each delay, we measure the magnitude of the Bloch vector projected in three equally spaced phase angles to extract a characteristic dephasing time $T_{2, \mathrm{echo}} = 0.943$~ms, from an exponential fit. Assuming photon shot noise as the dominant source of dephasing, we derive an average of $4 \times 10^{-3}$ residue photons in the readout cavity (see Appendix~\ref{appendix:photon_noise}).

In \autoref{fig:t1_t2_sweetspot}b, we show the temporal fluctuations of $T_{1}$ in a time span of $\sim 60$ hours for qubit G and H as labelled in \autoref{tab:coherence_metrics}, with the minimum qubit frequency $f_{01} = 197$ and $153$~MHz respectively. Each energy relaxation curve is fitted with a single exponential decay function. 
We obtain an average $T_1$ of $1.07\pm 0.18$~ms and $0.87\pm 0.30$~ms for these two devices, while a $T_{1}$ variation between its minimum and maximum values of 2-3 times is observed. Such behavior is consistent with the temporal fluctuation of the two level system (TLS) defects~\cite{klimov2018fluctuations, burnett2019decoherence}.  

\section{Dielectric loss and flux noise}

\begin{figure*}[bt]
    \includegraphics[width = 1\textwidth]{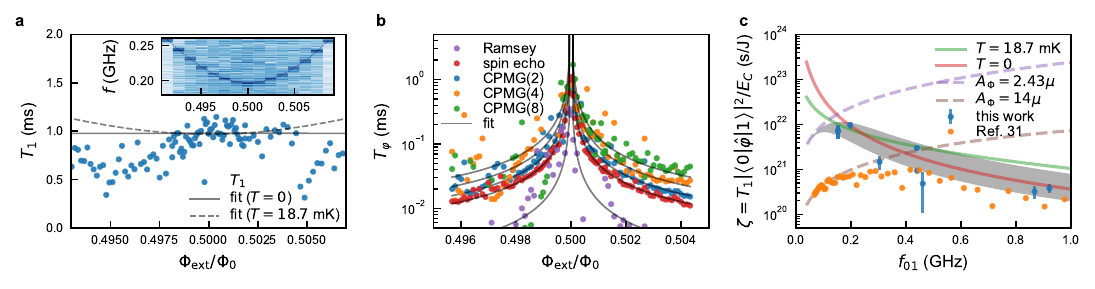}
    \caption{
    $\bf{a}$ Relaxation times $T_{1}$ versus $\Phi_{\mathrm{ext}}$ for qubit G, with a minimum frequency ($f_{01}$) of 197 MHz. Solid and dashed lines represent the dielectric loss model (\autoref{eq:dielectric_loss}) predictions at $T=0$ and $T=18.7$ mK, respectively. Inset: Qubit spectrum near flux frustration point.
    $\bf{b}$ Pure dephasing times $T_{\varphi}$ versus $\Phi_{\mathrm{ext}}$ for the same qubit, using free induction (Ramsey) and various dynamical decoupling sequences (spin echo, CPMG(2), CPMG(4), CPMG(8)). The black lines indicate dephasing times due to flux noise derived from a joint fit of all data. $\bf{c}$ $\zeta$ versus $f_{01}$ for our qubits in \autoref{tab:coherence_metrics} (blue dots) and the $E_J$-tunable qubit from Ref.~\cite{Sun2023} (orange dots). Green and red solid lines represent $\zeta(f_{01})$ calculated using the dielectric loss model (\autoref{eq:dielectric_loss}) at $T=18.7$~mK and $T=0$, respectively. Predictions using the flux noise model with different $A_{\Phi}$ are shown as dashed lines. The grey shaded area signifies predicted $\zeta(f_{01})$ combining dielectric loss ($\tan\delta_C \in (1.21, 4.93) \times 10^{-6}$ at $T=0$) and flux noise at $A_\Phi = 2.43~\mu\Phi_{0}/\sqrt{\text{Hz}}$.
    } 
    \label{fig:t1_t2_vs_phi}
\end{figure*}

We measure qubit coherence times across various external flux positions, $\Phiext$, to extract the dielectric loss tangent, $\tan\delta_{C}$, and $1/f$ flux noise amplitude, $A_{\Phi}$. Followed by high fidelity qubit initialization to its ground state~\cite{wang2024efficient}, we employ a standard single $\pi$ pulse to prepare the excited state for $T_1$ measurements and utilize either Ramsey or dynamical decoupling sequences for $T_2$ measurements. The experiments are conducted through microwave-driven qubit rotations lasting 40~ns each and dispersive readouts of 2~$\mu$s duration, all at a specific flux position referred to as the idle position $\Phi_\text{idle}$. During qubit decay, we adiabatically shift the external flux to various points, allowing us to sample the decoherence process in relation to $\Phiext$ efficiently, without recalibrating the qubit operations at each $\Phiext$. 

In \autoref{fig:t1_t2_vs_phi}a and b, we present the measured $T_{1}$ and pure dephasing $T_{\varphi}$ times of qubit G as functions of $\Phiext$, respectively.
Each data point for $T_{1}$ is extracted from fitting the relaxation data with an exponential decay curve at a specific flux position. 
The $T_{1}$ versus $\Phiext$ data exhibit a plateau near the flux frustration position and demonstrate suppression at $\Phiext = \Phi_0/2\pm 0.005\Phi_0$, likely due to interaction with a TLS at the corresponding qubit frequency. This flux dependence of $T_1$ precludes inductive-like loss channels~\cite{vool2014non}.
This data is subsequently fitted using two variants of the dielectric loss model to determine the dielectric loss tangent, $\tan\delta_{C}$. One variant adheres to the widely used phenomenological dielectric loss model~\cite{nguyen2019high, zhang2021universal}:
\begin{equation}
      \frac{1}{T_{1}^{\mathrm{diel}}} = \frac{\hbar \omega_{01}^2}{4E_C}\tan\delta_C \abs{\langle 0|\hat \varphi|1\rangle}^2\coth\left(\frac{\hbar\omega_{01}}{2k_B T}\right), \label{eq:dielectric_loss}
\end{equation}
dependent on the qubit's finite effective temperature $T$. 
The alternative model posits a bath of TLS weakly interacting with the qubit, demonstrating temperature independent loss by taking the saturation effect of the TLS at finite temperature into account~\cite{Sun2023}. This model coincides with \autoref{eq:dielectric_loss} at $T \rightarrow 0$, where the $\coth$ term converges to unity. The choice of model influences the derived loss tangent $\tan\delta_{C}$, typically yielding a higher value for the TLS bath model when the qubit energy $\hbar\omega_{01} \lesssim k_B T$.
The qubit parameters such as the matrix element $\langle 0|\hat \varphi|1\rangle$, the charging energy $E_C$, and the qubit frequency $\omega_{01} = 2\pi f_{01}$ are independently obtained from the qubit spectroscopy, where a representative spectrum of the corresponding qubit around the flux frustration position ($\Phiext = \Phi_0/2$) are shown in the inset of the \autoref{fig:t1_t2_vs_phi}a. 
We observe occasional suppression of $T_{1}$ at $\Phiext = \Phi_0/2$, suggestive of a TLS drifting into resonance with the qubit, when the qubit is idled at the same specific external flux position ($\Phi_\text{idle} = \Phi_0/2$). This suppression, however, can be mitigated by idling the qubit at $\Phi_{\text{idle}} \neq \Phi_0/2$ (see Appendix~\ref{appendix:measurement_protocols} for details). For instance, the data in \autoref{fig:t1_t2_vs_phi}a is collected with the qubit idled at $\Phi_{\text{idle}} = 0.505\Phi_0$. This behavior might stem from alterations in the TLS environment triggered by qubit dynamics~\cite{spiecker2022a}, presenting an intriguing avenue for future research.

In \autoref{fig:t1_t2_vs_phi}b, we present the typical qubit pure dephasing times $T_{\varphi}$ as a function of $\Phiext$, obtained from different dynamical decoupling schemes that utilize Ramsey, spin echo, CPMG($N$)~\cite{carr1954effects, meiboom1958modified} pulse sequences (with $N = 2, 4, 8$ representing the number of $\pi$ pulses).
Each data point is the result of fitting a Gaussian decay to the dephasing~\cite{yoshihara2006decoherence, bylander_2011_noise, yan2012spectroscopy}. 
At the flux frustration point ($\Phiext = \Phi_{0}/2$), where the qubit exhibits first-order insensitivity to flux noise, we record the maximum $T_\varphi$. This dephasing time $T_\varphi$ decreases monotonically with deviation from this optimal position, a behavior typical of flux-type qubits~\cite{yoshihara2006decoherence}.
The complete $T_\varphi$ versus $\Phi_{\mathrm{ext}}$ dataset, including Ramsey, spin echo, and CPMG($N$) sequences, is then fitted to a $1/f$ flux noise model~\cite{bylander_2011_noise}, \textit{i.e.} $S_{\Phi}(\omega) =  A_{\Phi}^{2} (2\pi \times 1~\text{Hz}/\omega)$ for the extraction of the flux noise amplitude $A_{\Phi}$ (see Appendix~\ref{appendix:data_analysis} for more details).

\begin{table*}
\begin{ruledtabular}
\begin{tabular}{ccccccccccc}
\makecell{ \\ Qubit} & \makecell{$f_{01}$ \\ (MHz)}  & \makecell{$E_C/h$ \\ (GHz)} & \makecell{$E_J/h$ \\ (GHz)} & \makecell{$E_L/h$ \\ (GHz)}  & \makecell{$T_1$ \\ (ms)} & \makecell{$T_{2, \mathrm{echo}}$ \\ (ms)} & \makecell{$T$ \\ (mK)} & \makecell{$\tan\delta_{C}$\\ ($T\neq 0$) \\ ($\times 10^{-6}$)} & \makecell{$\tan\delta_{C}$\\ ($T = 0$) \\ ($\times 10^{-6}$)} & \makecell{$A_{\Phi}$ \\ ($\mu\Phi_{0}/\sqrt{\text{Hz}}$)}  \\
\colrule
A & 921 & 1.369 & 2.758 & 0.585 &  0.072$^{*}$ & 0.098$^{*}$  & 16.2 & 1.49 &1.65 & 2.06  \\
B & 867 & 1.358 & 2.836 & 0.573 & 0.059$^{*}$ & 0.095$^{*}$ & 18.6 &  1.05 & 1.21 & 1.54 \\
C & 461 & 1.263 & 6.654 & 1.406 &  0.076$^{*}$ & 0.015$^{*}$ & 17.6 & 2.30 & 2.64  & 1.86  \\ 
D & 440 & 1.310 & 4.175 & 0.572 & 0.129 & 0.217 & 12.4 & 2.85 & 3.57 & 2.57 \\
E & 439 & 1.317 & 4.168 & 0.552 &  0.413 & 0.345  & 11.8 & 1.45 & 1.84 & 2.35 \\
F & 304 & 1.205 & 7.604 & 1.497 &  0.204$^{*}$ & 0.111$^{*}$ & 17.5 & 2.95  & 4.93 & 1.42 \\
G & 197 & 1.212 & 5.315 & 0.547 &  1.070 & 0.943 & 18.7 & 0.79 & 2.64 & 2.43 \\
H & 153 & 1.441 & 7.072 & 0.535 & 0.873 & 0.355 &  - & - & - & - \\ 
\end{tabular}
\end{ruledtabular}
\caption{\label{tab:coherence_metrics}Summary of device parameters and coherence metrics. $T_{1}$ and $T_{2, \mathrm{echo}}$ values, are obtained by fitting relaxation and dephasing to exponential decays at the flux frustration position. Asterisks (*) highlight observations of non-exponential $T_{1}$ decay. For qubit G, $\tan\delta_{C}$ and $A_{\Phi}$ are extracted from coherence time data as a function of $\Phiext$, collected with $\Phi_\text{idle} = 0.505\Phi_0$. Coherence times versus $\Phi_{\text{ext}}$ for qubit H were not collected.}
\end{table*}

\autoref{tab:coherence_metrics} presents the coherence metrics of various qubit devices, all fabricated using the same overlap junction process, characterized by their $T_{1}$ and $T_{2, \mathrm{echo}}$ values at $\Phiext = \Phi_0/2$, as well as extracted dielectric loss and flux noise levels. The qubits, denoted by asterisks, showed non-exponential decay in energy relaxation that led to reduced and fluctuating $T_1$ times. Notably, this phenomenon primarily emerges at specific $\Phiext$, thereby marginally influencing the reported dielectric loss, which is extracted by fitting the comprehensive $T_{1}$ versus $\Phiext$ data. This behavior aligns with qubit interactions with near-resonant TLS~\cite{klimov2018fluctuations,spiecker2022a,Sun2023}. Across a broad spectrum of qubit parameters resulting different $f_{01}$, we observed dielectric loss tangents, $\tan\delta_{C} \in (1.2, 5.0)\times 10^{-6}$, and $1/f$ flux noise amplitudes, $A_\Phi \in (1.4, 2.6)~\mu\Phi_0/\sqrt{\text{Hz}}$. These findings attest to the effectiveness and consistency of our overlap junction process in achieving low noise levels in fluxonium qubits.

\section{Discussion}

To account for various qubit parameters, we normalize $T_1$ to a quantity $\zeta = T_{1} \abs{\langle 0|\hat \varphi| 1\rangle}^2 / E_{C}$ to reflect its general frequency dependent part as motivated from the loss model in \autoref{eq:dielectric_loss}. In \autoref{fig:t1_t2_vs_phi}c, we plot $\zeta$ as a function of qubit frequency $f_{01}$ at the flux frustration position for all the measured qubits made from overlap junctions alongside with data from a $E_J$-tunable fluxonium fabricated with a bridgeless angle evaporation process from Ref.~\cite{Sun2023}. 

The frequency dependence of coherence from the presented overlap junction qubits is different from that found in Ref.~\cite{Sun2023}, where the qubit relaxation is found limited by flux noise with an excessive $1/f$ noise amplitude $A_{\Phi} = 14~\mu \Phi_0/\sqrt{\text{Hz}}$ when $f_{01} \lesssim 400$~MHz. 
In addition, we collect coherence data from fluxonium qubits with a Manhattan-style JJA design, fabricated with the identical overlap junction process (see Appendix~\ref{appendix:additional_data}). The extracted flux noise amplitude $A_\Phi \in (3.7, 8.2)~\mu\Phi_0/\sqrt{\text{Hz}}$ from those devices is appreciably higher than that from the qubits with the compact JJA design presented in the main text.
We attribute the flux noise reduction to the formation of compact qubit loop with reduced susceptibility to the magnetic spin defects located at the interfaces of the circuits~\cite{lanting2009geometrical, anton2013magnetic, braumuller2020characterizing}. Owning to this flux noise reduction, we observe a trend of increasing in $T_{1}$ at the flux frustration position with a reduction of the qubit frequency $f_{01}$, providing evidences to refine the dielectric loss model for fluxonium.
 
Earlier research, which reports relaxation in devices across a broad spectrum of qubit frequencies~\cite{nguyen2019high, Sun2023}, employs \autoref{eq:dielectric_loss} with a frequency dependent loss tangent $\tan\delta_C \propto \omega^\epsilon$, $\epsilon > 0$ as an empirical model, originated from qubit interactions with a bosonic bath at finite temperature $T$~\cite{devoret1997quantum}. This model, which sees $\tan\delta_C$ increase with frequency, is consistent with observations of high coherence fluxonium~\cite{somoroff2021millisecond, ding2023highfidelity}, where the dielectric loss deduced from the $\ket{2}\rightarrow \ket{1}$ relaxation (with $\sim 5$~GHz transition frequencies) is substantially higher than that within the computational space (with $\sim 200$~MHz transition frequencies).
However, it treats $\epsilon$ as a free variable and does not elucidate the loss's microscopic origins.

Conversely, recent studies~\cite{somoroff2021millisecond, spiecker2022a, Sun2023} point to material defects, or TLS, as a primary relaxation source, akin to transmons.
This perspective posits that dielectric loss from TLS interactions should be temperature-independent, thus rendering $T=0$ in \autoref{eq:dielectric_loss}. Consequently, the TLS model predicts a more rapid increase in $T_1$ with decreasing qubit frequency ($\zeta \propto 1/f_{01}^{2}$), eliminating the need of frequency-dependent loss tangent in the empirical model.

We plot both dielectric loss models in \autoref{fig:t1_t2_vs_phi}c. 
A model incorporating TLS dielectric loss ($\tan\delta_C \in (1.21, 4.93) \times 10^{-6}$) and $1/f$ flux noise ($A_\Phi = 2.43~\mu\Phi_0/\sqrt{\text{Hz}}$), represented by the grey shaded area, aligns well with our data and simplifies the explanation of fluxonium energy relaxation over a broad frequency range. Our findings indicate that reducing dielectric loss and flux noise is essential for extending fluxonium coherence times beyond the 1-millisecond level.

\section{Conclusion}

To summarize, we fabricated and measured fluxonium qubits with millisecond coherence times based on overlap junction process. The process eliminates the need for angle evaporation, and enables almost 100\% qubit yield. On a 2-inch wafer, we achieved relative standard deviations below 5\% and 2\% for phase slip junctions and junction arrays, respectively, with typical junction sizes for fluxonium. The achieved dielectric loss and flux noise levels obtained from multiple devices are consistent with some of the most coherent superconducting qubits to date~\cite{nguyen2019high,somoroff2021millisecond, ding2023highfidelity}. 

Our data indicates that the quality of qubits is not necessarily compromised by breaking the vacuum during the material formation for Josephson junctions. Our work thus suggests a path toward a CMOS-compatible fabrication process that can contribute to the scaling-up of high coherence fluxonium processors. This addresses the manufacturing challenge of integrating approximately 100 junctions per qubit and showcases the method's versatility in accommodating a wide range of junction geometries. Additionally, our approach opens avenues for material system optimization in Josephson junctions, moving beyond the conventional Al/AlO$_x$/Al structure.

\begin{acknowledgments}
The experimental part of this work was conducted at DAMO Quantum Laboratory, Alibaba Group. At the time of the laboratory's announced closure in November 2023, all authors were affiliated with DAMO Quantum Laboratory. We acknowledge the support from Guangdong Provincial Quantum Science Strategic Initiative (Grant No. GDZX2407001). Feng W. and H-H.Z. are supported by Zhongguancun Laboratory.
\end{acknowledgments}

\appendix

\section{Fabrication process}\label{appendix:fabrication}
Devices are fabricated on a 2-inch c-cut sapphire substrate. A 180~nm thick tantalum film is sputtered onto the substrate as the base metal layer for superconducting circuits, and the wafer is spin-coated with 1~$\mu m$ S1813 photoresist for following patterning. All circuits except for Josephson junctions are patterned by direct laser writing system, and the development is performed in MIF-319 followed by DI water rinse. The tantalum film is then etched in a RIE system using BCl3/Ar mixed gas to achieve decent sidewall profile. The photoresist is stripped in a sequence of baths of EKC270, acetone and IPA. The adoption of EKC270 has been proven to be very effective at removing etch residuals through resonator quality factor measurements. The wafer is then soaked in piranha solution at room temperature for 10~min as a final clean before junction formation. 

\begin{figure}
    \includegraphics[width = 1\columnwidth]{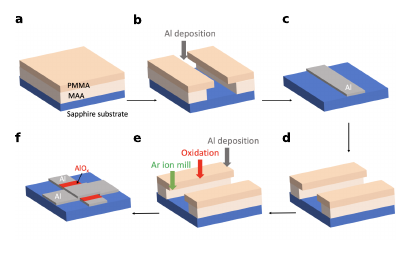}
    \caption{Overlap junction process flow. \textbf{a-c} Bottom Al electrode patterning, deposition and lift-off; \textbf{d} Patterning for the 2nd part of junction process; \textbf{e} Sequence of Ar ion milling, static oxidation and Al deposition; \textbf{f} Final lift-off to complete junction formation.}
    \label{fig:overlap_process_flow}
\end{figure}

With all circuits formed on the tantalum base layer, the Al/AlO$_x$/Al Josephson junctions (JJ) and Josephson junction arrays (JJA) are then fabricated using an overlap junction process with lift-off technique, which requires two times of e-beam lithography patterning, shown in \autoref{fig:overlap_process_flow}. The wafer is spin-coated with a bilayer e-beam resist consisting of 600~nm PMMA A7 and 350~nm MAA EL9, and patterned in JEOL JBX8100FS e-beam lithography system. The development is performed in a bath of MIBK/IPA 1:3 for 2~min followed by 2~min of IPA rinse. The 1st layer of 60~nm Al film is e-beam evaporated onto the wafer in a high vacuum chamber, and the lift-off is performed in NMP at 70~$^{\circ}$C for 1 hour. The 2nd e-beam lithography is then performed in the same way to pattern the top electrode. The patterned wafer is loaded into the high vacuum chamber again for in situ oxidation and top electrode deposition. Prior to the oxidation, an Ar ion milling process is implemented to remove the native AlO$_x$ on bottom Al electrode. This process is crucial to the formation of high quality barrier AlO$_x$ layer in Josephson junctions, and will largely determine the qubit performance. An RF ion source is used to perform the Ar milling, and a two-step process is developed to achieve clean Al surface without any damage of e-beam resist. As described in Section II, the Ar ion milling process thins down the bottom Al electrode from 60~nm to 45~nm. The barrier AlO$_x$ layer is formed by room temperature static oxidation at 6~Torr O$_2$ pressure for 40~min, and the layer thickness is in the range of 1.5-2~nm from TEM characterization. Finally, 100~nm Al is e-beam evaporated onto the wafer as the top electrode, followed by an in situ oxidation to form a passivation layer. 

\section{Uniformity across 4-inch wafer}\label{appendix:4in}
\begin{figure}
    \includegraphics[width = 1\columnwidth]{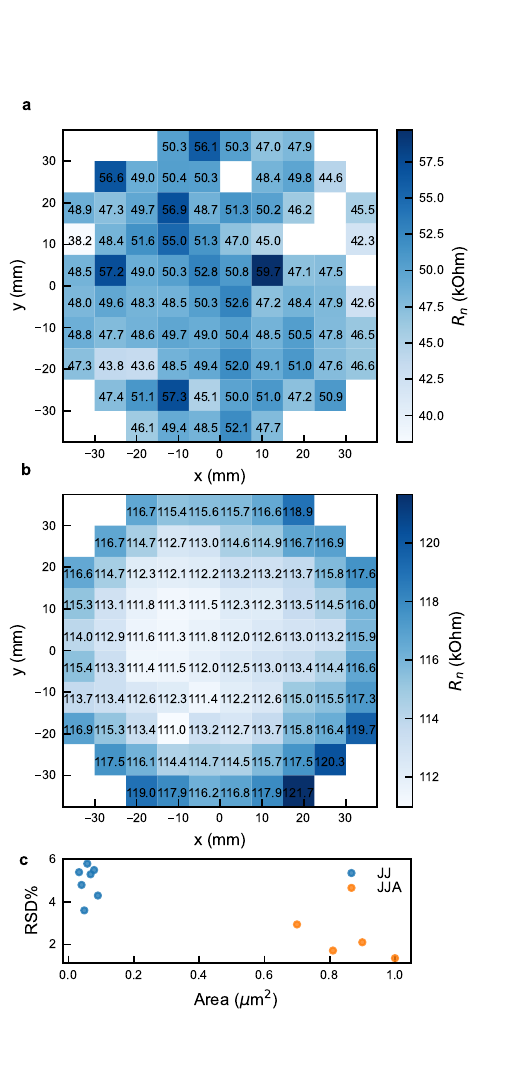}
    \caption{\textbf{a} and \textbf{b} Wafer maps showing $R_{n}$ for small Josephson junctions (JJ) and Josephson junction arrays (JJA) respectively. \textbf{c} Relative standard deviation (RSD\%) of $R_{n}$ across the wafer for both JJ and JJA.}
    \label{fig:4_in_rsd}
\end{figure}

To validate the scalability of the overlap junction process, we test it on 4-inch wafers. 
For examining junction yield and parameter variation, we replicate the test structures used on 2-inch wafers, utilizing the room temperature resistance ($R_{n}$) of the small junctions (JJ) and JJ arrays (JJA) as the primary criteria. We arranged 88 test dies on a 4-inch wafer, yielding 1408 $R_{n}$ measurements across different JJ/JJA dimensions. With the exception of some dies affected by known photolithography issues, nearly all JJ/JJAs exhibited $R_{n}$ values within the expected range, as illustrated in \autoref{fig:4_in_rsd}a and b, confirming a high junction yield. As depicted in \autoref{fig:4_in_rsd}c, the relative standard deviation (RSD\%) for JJ and JJA varies between 1.3\% and 6\%, depending on junction size. The consistency of RSD\% values between 4-inch and 2-inch wafers is an encouraging indicator of our ability to scale the overlap junction process to larger wafers without significant trade-offs in process variation.

\section{Cryogenic experimental setup}
\label{appendix:experimental_setup}
\begin{figure}
    \includegraphics[width = 1\columnwidth]{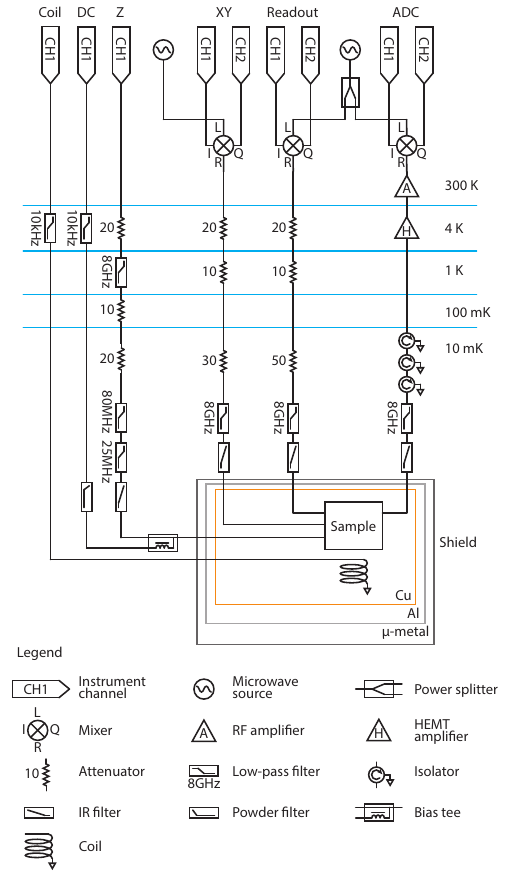}
    \caption{Schematics of the experimental setup. The sample is mounted in the cold finger with three layers of shielding to minimize the stray magnetic field at the mixing chamber of the dilution refrigerator. The input lines consist of DC and Z controls for tuning the external flux, XY line for driving qubits and Readout line for generating readout tone. Additionally, we also supply a coil as an alternative DC line to avoid heating when the sample loop is small. The XY and the Readout drives are supplied with 60~dB and 80~dB attenuation in the DR with a pulse duration of $40~\mathrm{ns}$ and $2~\mathrm{\mu s}$, respectively. An extra 20~dB attenuation is applied to the Readout drive at room temperature, which is not shown. The output signal is amplified at both 4~K and 300~K stages.}
    \label{fig:meas_setup}
\end{figure}

The qubit characterization experiment is performed in a dilution refrigerator at a base temperature below 10~mK. 
The setup incorporates wiring and room temperature electronics akin to those in Ref.~\cite{bao2022fluxonium}, albeit with adjustments: the flux control line (Z, as depicted in \autoref{fig:meas_setup}) includes 50~dB of total attenuation, and with 25~MHz and 80~MHz low pass filters in series, adapted to low-frequency qubits.
Accordingly, we use nominally 15~ns rise/fall time for the flux pulse. The flux pulse distortion is corrected with methods detailed in Ref.~\cite{Sun2023}. A superconducting coil, mounted on the sample holder, generates flux bias at the idle position, preventing heating due to large DC in the on-chip flux line. Extra room temperature 20~dB attenuation at the readout input is used, which is not shown in \autoref{fig:meas_setup}.

\section{Pulse measurement protocols} \label{appendix:measurement_protocols}

\begin{figure}
    \includegraphics[width = 1\columnwidth]{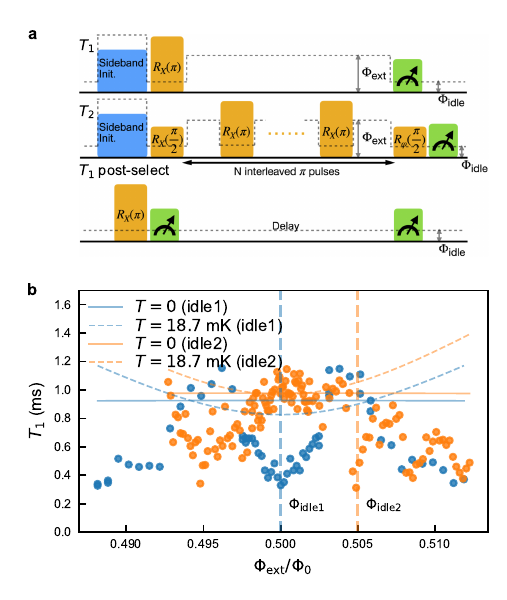}
    \caption{$\bf{a}$ Protocol for measuring $T_{1}$ and $T_{2}$ via fast flux pulses, and that for $T_{1}$ with static flux bias through heralded state preparation. For $T_{2}$ CPMG($N$) pulse sequence, there are $N$ $\pi$ pulses interleaved in between two $\pi/2$ pulses by convention. Dashed lines indicate the applied flux profiles in the sequence. $\bf{b}$ Dependence of $T_1$ on external flux $\Phiext$ at two idle positions, marked by vertical dashed lines. Dielectric loss models are shown as lines.}
    \label{fig:t1_different_idle_point}
\end{figure}

We measure coherence times $T_{1}$ and $T_{2}$ against the external flux employing methods akin to those in Ref.~\cite{barends2013coherent,bao2022fluxonium,Sun2023}, as illustrated in \autoref{fig:t1_different_idle_point}a. 
We bias the qubit towards its flux frustration point, at a static flux position $\Phi_{\text{idle}}$, facilitating qubit gates and readout operations. At the beginning of each pulse sequence, the qubit is initialized to its ground state through a sideband transition enabled by shifting it from its flux frustration position~\cite{wang2024efficient}, indicated by blue blocks and elevated flux pulse amplitude.
To explore coherence times across different flux values, we adiabatically shift the qubit to $\Phi_{\text{ext}}$ with rapid flux pulses during measurements. For $T_{2}$ measurement, we ascertain the Bloch vector's decay in the XY plane using three $\pi/2$ pulses at phases $\varphi = 0$, $\pi/3$, and $2\pi/3$, immediately preceding readout. This tri-phase approach facilitates the quantification of dephasing, visualized as a decaying Bloch vector amplitude. This tri-phase approach allows us to extract the decaying envelope directly, simplifying the fitting model and improving measurement accuracy. In the main text, Fig.~3(a) presents the $T_{2}$ dephasing profile and its corresponding fit, both elevated by 0.5 to enhance clarity.

In the $T_{1}$ measurement at the flux frustration position, where flux pulses are absent, we initialize the qubit through projective readout and herald its state~\cite{johnson_2012_heralded, ding2023highfidelity}. The qubit state is first identified through a readout, acting as the initialization step. Following a relaxation time delay for the qubit, we conduct a secondary readout. This complete sequence occurs every 5~ms and was performed 20,000 times for each delay setting in the data present in Fig.~3a. For individual delays, we compute the probability of the qubit being in the excited state $\ket{1}$, labelled $P_{1, \ket{1}}$, after the initial readout indicates a state of $\ket{1}$. Conversely, when the initial readout indicates a state of $\ket{0}$, the corresponding excited state probability is marked as $P_{1, \ket{0}}$.

In the $T_1$ versus $\Phi_\text{ext}$ measurement, we occasionally noted a $T_1$ dip that depends on the qubit's idle position, $\Phi_\text{idle}$, as illustrated in \autoref{fig:t1_different_idle_point}b. To extract the dielectric loss values for all qubits listed in Table~1 and \autoref{tab:coherence_metrics_additional}, we typically idle the qubit at $\Phi_\text{idle} = \Phi_0/2$ using a DC flux bias and shift it to the desired $\Phi_\text{ext}$ using a flux pulse for decoherence characterization across $\Phi_\text{ext}$. For qubit G, we observed a $T_1$ dip at $\Phi_\text{ext} = \Phi_0/2$ when idled at $\Phi_\text{idle} = \Phi_0/2$ (blue points). Idling the qubit at $\Phi_\text{idle} = 0.505\Phi_0$ eliminated this dip and introduced a new dip at the new idle position (orange points), hinting at a $T_1$ variation tied to changes in the TLS environment due to qubit dynamics. A thorough investigation might clarify this behavior and mitigate TLS poisoning. We compared dielectric loss models using data from idle positions at $\Phi_{\text{idle1}} = \Phi_0/2$ and $\Phi_{\text{idle2}} = 0.505\Phi_0$ in \autoref{fig:t1_different_idle_point}b, obtaining similar loss tangent, $\tan\delta_C$, for both. To prevent overfitting to data poorly described by the model, we biased our analysis towards data points with higher $T_1$ values, discussed further in the subsequent section.

\section{Non-exponential $T_1$ decays}
\label{appendix:nonexponential_decay}

\begin{figure}
    \includegraphics[width = 1\columnwidth]{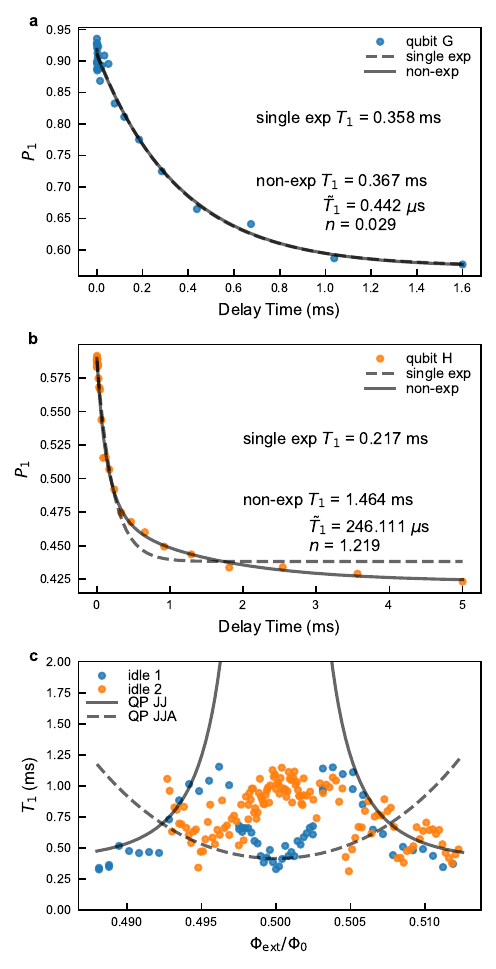}
    \caption{\textbf{a, b} The worst observed $T_1$ decay data for qubits G and H at $\Phiext = \Phi_\text{idle} = \Phi_0/2$, analyzed using single exponential and non-exponential decay models. \textbf{c} $T_1$ as a function of $\Phi_\text{ext}$ at two idle positions, as shown in \autoref{fig:t1_different_idle_point}b, compared with two loss models: quasiparticle tunneling across the phase-slip junction (solid line) and in the JJA (dashed line).}
    \label{fig:nonexponential}
\end{figure}

In the $T_1$ measurement, we sometimes observe non-exponential decay instead of the expected single exponential decay, as discussed in the main text. This phenomenon was initially attributed to quasiparticle poisoning~\cite{pop2014coherent, gustavsson2016suppressing}, but it was later found to also align with relaxation due to fluctuating two-level defects~\cite{spiecker2022a}. For instance, when monitoring the temporal fluctuation of $T_1$ for qubit H, as shown in \autoref{fig:t1_t2_sweetspot}b, the relaxation profile from the worst few $T_1$ data points fits poorly to a single exponential function $P_{1}(t) = a\exp(-t/T_{1})+b$. Instead, the data fits better to a non-exponential decay function, 
\begin{equation}
    P_{1}(t) = a e^{[n(\exp(-t/{\tilde T_{1}}-1)]} e^{(-t/T_{1})} + b
\end{equation}
with $n = 1.219$, $\tilde T_1 = 0.246$ ms, and $T_{1} = 1.464$ ms, as shown in \autoref{fig:nonexponential}b. In our noise analysis, data exhibiting such non-exponential decays are fitted with a single exponential model, resulting in $T_1$ values that are generally lower than those predicted by dielectric loss and flux noise models.

In the $T_1$ dip observed at the qubit's idle position—the blue data point corresponding to $\Phiext = \Phi_\text{idle} = \Phi_0/2$ in \autoref{fig:t1_different_idle_point}b—there is no clear evidence of non-exponential decay. \autoref{fig:nonexponential}a displays the $T_1$ trace fitted with both single and non-exponential decay models.

One possible cause of the non-exponential decay is relaxation due to quasiparticle tunneling, with the number of responsible quasiparticles fluctuating according to a Poisson distribution. We explore this possibility by analyzing the $T_1$ vs. $\Phi_{\text{ext}}$ data using two models~\cite{glazman2021bogoliubov}: quasiparticle loss due to tunneling across the phase-slip JJ
\begin{equation}
      {\Gamma_{1,\text{QP}}^{\mathrm{JJ}}} =  \abs{\langle 0|\sin\frac{\hat \varphi}{2}|1\rangle}^2 \frac{8E_J}{\hbar\pi}x_{\mathrm{QP}}\sqrt{\frac{2\Delta}{\hbar\omega_{01}}}\left(1+e^{-\frac{\hbar\omega_{01}}{k_{B}T}}\right), \label{eq:qp_loss_jj}
\end{equation} 
and in the JJA
\begin{equation}
      {\Gamma_{1,\mathrm{QP}}^{\mathrm{JJA}}} =  \abs{\langle 0|\frac{\hat{ \varphi}}{2}|1\rangle}^2 \frac{8E_L}{\hbar\pi}x_{\mathrm{QP}}\sqrt{\frac{2\Delta}{\hbar\omega_{01}}}\left(1+e^{-\frac{\hbar\omega_{01}}{k_{B} T}}\right), \label{eq:qp_loss_jja}
\end{equation} 
Using the characterized qubit parameters, the aluminum superconducting gap $\Delta$, and appropriately chosen average quasiparticle density $x_{\text{QP}}$, which may differ between the JJ and JJA, we find that neither model adequately describes the $T_1$ data around the flux frustration position, as shown in \autoref{fig:nonexponential}c. The $T_1$ dip observed near $\Phiext = \Phi_0/2$ corresponds to a spectral linewidth of less than 20~MHz, which is more consistent with relaxation due to a two-level system (TLS) with a random frequency distribution.

\section{Noise spectral density data processing}
\label{appendix:data_analysis}

The $T_1$/$T_2$ decoherence measurement data are processed similarly to Ref.~\cite{Sun2023}, yielding the dielectric loss tangent and flux noise amplitude. Given the likelihood of TLS defects with strong qubit coupling, the qubit's $T_{1}$ could drop below the dielectric loss model's projection (Eq.~1) at specific flux locations due to resonances with TLS. This contrasts with the model's expected smooth $T_1$ variation with $\Phiext$, based on averaging over weakly coupled charge defects in the material~\cite{martinis2005decoherence}. To accommodate this, our fitting approach for determining $\tan\delta_{C}$ prioritizes positive $T_1$ deviations, aiming for the model to approximate the upper limit of the $T_1$ data. Consequently, the inferred dielectric loss tangent should be considered as presenting a lower bound estimate, excluding effects from resonances with strongly coupled TLSs.

In the processing of $T_2$ data, we calculate the CPMG pulse sequence's filter functions with $N$ pulses~\cite{bylander_2011_noise} and subsequently compute the decay function $f_N(t)$ for data fitting. We assume the noise spectral density of the flux noise to be a combination of a $1/f$ component and a white noise component. The latter, unaffected by the filter function, consistently causes an $N$-independent exponential decay, $f_{\text{white}}(t)$, in all $T_2$ measurements.

The dephasing function $f_N(t)$ due to $1/f$ flux noise takes a Gaussian shape, $f_N(t)=\exp[-(tD A_\Phi u(N))^2]$, where $A_\Phi$ represents the flux noise amplitude, $D = \partial \omega_{01}/\partial \Phiext$ denotes the derivative of frequency to the external flux (derivable from fluxonium models), and $u(N)$ is a sequence-dependent coefficient determined directly from filter functions~\cite{ithier2005decoherence}. Specifically, for spin-echo, $u(1)=\sqrt{\ln 2}$, and $u(N)$ decreases with increasing pulse numbers. Although not distinctly Gaussian, the decay function from Ramsey experiments can be closely approximated as Gaussian with a low-frequency cutoff, adopted here as 1~Hz.

The collective $T_2$ measurement is described by
\begin{equation}
    \chi(t) = a(f_{T_1}(t/2)f_{\text{white}}(t)f_N(t)), \label{eq:cpmg_decay}
\end{equation}
in which $a$ denotes $\chi(t)$ at $t=0$ and $f_{T_1}(t)$ is the $T_1$ decay extracted from the $T_1$ measurement at the relevant $\Phiext$. By fitting decay curves across various $N$ and $\Phiext$, we can parameterize the fits using merely 2 variables associated with the noise model, $A_{\Phi}$ and $A_{\text{white}}$.

\begin{figure}
    \includegraphics[width = 1\columnwidth]{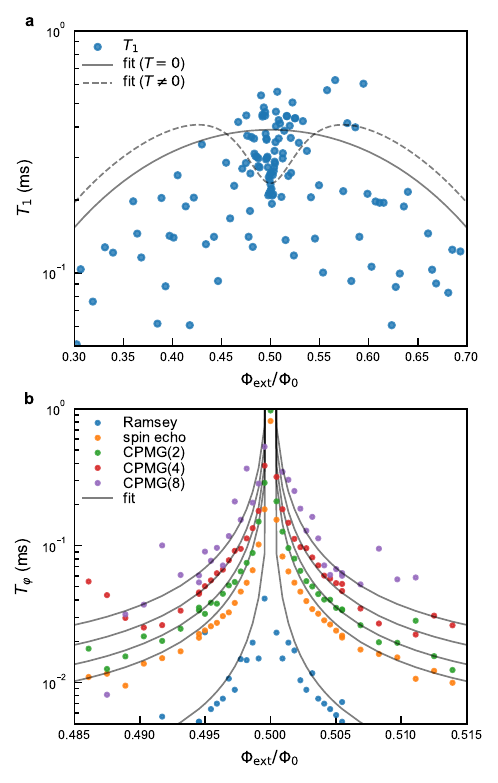}
    \caption{$\bf{a}$ $T_1$ versus $\Phiext$ for qubit E. Solid and dashed lines represent the dielectric loss model predictions at $T = 0$ and $T = 11.8$~mK, respectively. $\bf{b}$ $T_\phi$ versus $\Phiext$ for the same qubit.}
    \label{fig:additionalt1t2phi}
\end{figure}

As a supplement, we present the coherence times as a function of $\Phi_{\mathrm{ext}}$ along with the corresponding fits used to extract the noise spectrum for qubit E in \autoref{fig:additionalt1t2phi}.

\section{Photon shot noise}\label{appendix:photon_noise}
For most of our devices, $T_{2}$ does not reach $2T_{1}$ even at the flux noise insensitive position ($\Phi_\text{ext} = \Phi_{0}/2$). We estimate that coherent quantum phase slips across the array junctions and second-order effects from flux noise independently contribute to a dephasing time limit of over 10~ms, based on realistic noise assumptions and our qubit parameters. Consequently, we attribute the observed dephasing primarily to photon fluctuations in the readout cavity. Given the qubit-cavity coupling $\chi$ and the cavity linewidth $\kappa$, the dephasing rate induced by photon shot noise is calculated to be
\begin{align}
    \Gamma^{\mathrm{ph}}_{\varphi} = \frac{\bar{n}\kappa}{1 + \kappa^{2}/\chi^{2}},
\end{align}
where $\bar{n}$ is the average photon number in the cavity. For qubit G, since $T_{1} = 1.07$ ms and $T_{2} = 0.943$ ms, we obtain $T_{\varphi} = 1.685$ ms. With the designed $\kappa/2\pi = 2.19$ MHz and $\chi/2\pi = 0.223$ MHz, we obtain an estimate of the cavity photon number $\bar{n} \approx 4\times 10^{-3}$. With the cavity frequency being $6.69$~GHz, we thus estimate the effective temperature of the readout cavity to be 59~mK.

\section{Qubits with Manhattan-style Josephson junction array} \label{appendix:additional_data}

\begin{table*}[t]
\begin{ruledtabular}
\begin{tabular}{ccccccccccc}
\makecell{ \\ Qubit} & \makecell{$f_{01}$ \\ (MHz)}  & \makecell{$E_C/h$ \\ (GHz)} & \makecell{$E_J/h$ \\ (GHz)} & \makecell{$E_L/h$ \\ (GHz)}&  \makecell{$T_1$ \\ (ms)} & \makecell{$T_{2, \mathrm{echo}}$ \\ (ms)} & \makecell{$T$ \\ (mK)} &  \makecell{$\tan\delta_{C}$\\ ($T\neq 0$) \\ ($\times 10^{-6}$)} & \makecell{$\tan\delta_{C}$\\ ($T = 0$) \\ ($\times 10^{-6}$)} & \makecell{$A_{\Phi}$ \\ ($\mu\Phi_{0}/\sqrt{\text{Hz}}$)}  \\
\colrule
S\_A & 609 & 1.403 & 3.716 & 0.570&  0.060 & 0.096 & 10.3 & 2.57 & 2.64 & 8.14 \\
S\_B & 593 & 1.391 & 3.792 & 0.583 &  0.107 & 0.122 & 14.3 & 1.62 & 2.01 & 4.07 \\
S\_C & 592 & 1.391 & 3.877 & 0.603&  0.129 & 0.167 & 12.8 & 2.30 & 2.52 & 3.72 \\ 
S\_D & 558 & 1.384 & 3.938 & 0.591&  0.112 & 0.196 &   19.3& 2.44 & 3.24 & 4.85 \\
S\_E & 550 & 1.385 & 3.949 & 0.574&  0.209 & 0.099 & 16.7 & 1.44 & 2.05 & 5.09 \\
S\_F & 477 & 1.370 & 4.274 & 0.598& 0.138 & 0.166 & 50.6 & 1.54 & 4.60 & 4.88 \\
\end{tabular}
\end{ruledtabular}
\caption{Summary of qubit coherence metrics with Manhattan-style JJA.}
\label{tab:coherence_metrics_additional}
\end{table*}

\begin{figure}
    \includegraphics[width = 1\columnwidth]{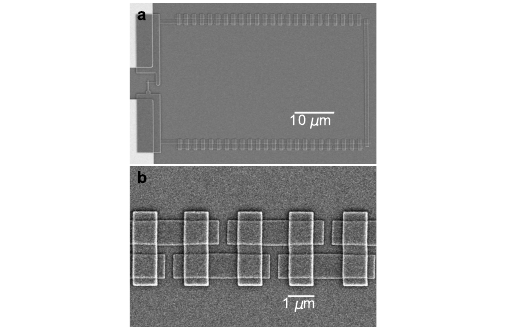}
    \caption{$\bf{a}$ Scanning electron microscopy (SEM) image of the qubit loop made with the Manhattan-style JJA. $\bf{b}$ Zoomed-in SEM image of the Manhattan-style JJA.}
    \label{fig:cross_jja}
\end{figure}

In \autoref{tab:coherence_metrics_additional}, we present coherence metrics for samples fabricated using the overlap technique in a Manhattan-style JJA geometry (\autoref{fig:cross_jja}), contrasting with the compact JJA geometry discussed in Figure~1 of the main text. Employing the same overlap technique, these samples exhibited a flux noise amplitude $A_{\Phi} > 3.7~\mu\Phi_{0}/\sqrt{\text{Hz}}$, as recorded in \autoref{tab:coherence_metrics_additional}. A detailed comparative study of the flux noise levels between compact and Manhattan-style JJA designs could shed light on the underlying mechanisms of flux noise. We hypothesize that the observed disparity in flux noise levels between the two designs might stem from their distinct geometries. In simpler systems like the SQUID, geometry has been demonstrated to influence flux noise, with larger cross-sectional areas and shorter perimeters correlating with reduced noise levels~\cite{braumuller2020characterizing}. Specifically, the compact JJA features wider loops ($\sim 2~\mu m$) compared to the Manhattan-style JJA ($\sim 1~\mu m$), likely leading to lower surface current densities and diminished sensitivity to surface spins. Additionally, assuming an equal number of Josephson Junctions (JJs), the compact configuration typically possesses a shorter loop perimeter, potentially accruing less flux noise than its Manhattan-style counterpart. Despite the complexity of JJ arrays, pursuing numerical and theoretical investigations into current distribution could provide further insights into these observations.

\bibliography{ref}

\end{document}